# How circular is the linear economy?
## Analysing circularity, resource flows and their relation to GDP

*Amir Rashid*

**Abstract**

The concept of Circular Economy (CE) has evolved significantly over the past decade, transitioning from a simple model of resource circulation to an increasingly complex and debated framework. While its primary objective remains the elimination of waste and pollution through regenerative processes, CE has encountered definitional ambiguities and criticisms. Despite these challenges, it has gained widespread support among researchers, policymakers, and businesses. This study critically examines the prevailing circularity metrics—such as the "circular material use rate" or "circularity" (e.g., Eurostat's 12% EU circularity rate)—and argues that such narrow definitions obscure the true potential of CE by excluding higher-value strategies like maintenance, repair, refurbishment, and remanufacturing. Through a mixed-methods analysis of global resource flows (e.g., 104 Gt input in 2020, with only 9% recycled), the study demonstrates how adjusting circularity calculations for non-recoverable materials (e.g., dissipative flows and stock additions) reveals a "real" circularity rate of 27%, far exceeding the "apparent" 9%. Yet even this higher rate translates to a mere 1.4% of global GDP, underscoring the limited economic impact of recycling-centric approaches. More significantly, the study identifies that 69% of economic value already derives from managing existing stocks, suggesting mainstream CE discourse has largely overlooked the most substantial circular practices already embedded in modern economies. The study therefore proposes a radical change in the assessment framework that, (a) replaces annual input-based metrics with economic value creation as the primary indicator, (b) incorporates stock utilization efficiency as a core circularity measure and (c) establishes new policy targets focused on value retention and reuse rather than mere material recovery. These findings necessitate a paradigm shift in circular economy strategy - from counting recycled materials to optimizing economic resilience through intelligent stock management and service-based value creation. In other words, this work calls for a paradigm shift in CE discourse, centering *economy* rather than *materials* to unlock feasible pathways toward sustainability.

**Introduction**

Popular concept of circular economy (CE), from its humble beginning to a storming rise has seen several phases of evolution in the past one decade. Simply meant to create an economy driven by circulation of resources to eliminate waste and pollution while regenerating the natural eco system, its evolution over the past decade has seen undue burden of complexity and confusion. From near drowning in the quagmire of hundreds of definitions[1,2,3] to facing criticism as peculiar as lack of clarity on CE's potential to create "religious equality"[4,5], circular economy appears to have survived the battle of its life. In this journey of survival, it met with both heroes and villains but has luckily created more friends than foes and now appears to enjoy an overwhelming support from kings and queens, fathers and grandfathers, influencers and lovers of circular economy and above all the circular economists. "Circular Economist" in this list is an intriguing title prompting to ponder over the known history of economy and the profession

---

[1] Julian Kirchherr et al., *Conceptualizing the circular economy: An analysis of 114 definitions*, Resources, Conservation and Recycling, Volume 127, December 2017, Pages 221-232
[2] Frank Figge et al., *Definitions of the circular economy: Circularity matters*, Ecological Economics, Volume 208, June 2023, 107823
[3] Julian Kirchherr et al., *Conceptualizing the Circular Economy (Revisited): An Analysis of 221 Definitions*, Resources, Conservation and Recycling, Volume 194, July 2023, 107001
[4] Alan Murray et al., *The Circular Economy: An Interdisciplinary Exploration of the Concept and Application in a Global Context*, Journal of Business Ethics, Volume 140, pages 369–380, (2017)
[5] Hervé Corvellec et al., *Critiques of the circular economy*, Journal of Industrial Ecology 2022;26:421–432

of economist. It is interesting to note that while linear economy preceded the emergence of its experts, circular economy has the honour of having its experts even before the concept itself is fully understood and established. It is difficult though to predict that this honour will eventually turn out be a blessing or otherwise. What is however clear is that ECONOMY in its true sense remains either absent or confusingly implicit in the discussion and analyses on circular economy. As we understand today, primary thesis of circular economy is based on the flows and stocks of resources in the socioeconomic system. It is nevertheless interesting that despite all important developments in the realm of circular economy and even linear economy so far, it is hard to find studies that can help one to understand the relationship between resource flows and stocks together with the economy they create today and have the potential to create if the principles of circular economy are used in transitioning from linear to circular.

This article is an attempt, to first understand the concept of circularity or more precisely the term "circularity rate" as used in high level discussions and communications especially in macroeconomic context, second to scrutinise how resource flows and stocks relate to economy as we know it today, third to explore how and to what extent circularity is in action in today's linear economy and fourth to understand what is the potential of circular economy principles in transforming the linear economy to a circular one. Primary intent of this analysis is to bring the term ECONOMY in focus, which is almost always ignored in discussions on circular economy.

Reflecting on the large body of knowledge developed and significant number of policy and funding initiatives taken in the past decade, as well as intensity of engagements by researchers, businesses, policy makers and governments alike, one can appreciate the transformative character of circular economy and awareness of it across the globe. However, despite significant efforts and resources dedicated to understanding and development of the field of CE, it remains vague if we are aiming at a full-scale replacement of the linear economy with a circular economy or only interested in exploiting the idea as an opportunity of estimated, $4-5 trillion[6] to merely add to existing size of the linear economy. Although in progressive circles of the society, it is emphasised as a necessary and viable alternative to the unsustainable linear economy of today, for conservative circles it may only be another lucrative business opportunity. Regardless of if it is an alternative to linear economy or another business opportunity, a clearer understanding of circular ECONOMY vis-à-vis linear economy is essential.

Since the practice of production, use and management of resources is fundamental to economy, following the flow of resources through the existing socioeconomic system can provide a valuable insight on the variety, magnitudes and final fate of the resources. Such an insight can then be used to understand associated financial or economic flows and in turn the economy itself, as it is practiced or popularly known through measures such as annual GDP, GDP per capita year, etc.

**Methodology:**

This work employs a mixed-method approach, combining quantitative analysis of global resource flows with a critical review of existing circularity metrics and their economic implications. The methodology is structured as data collection and framework, critical review of circularity metric, resource flow analysis, economic value mapping, scenario analysis and policy implications.

**Critical review:**

Before delving deeper into the resource flows and their relationship with economy, understanding of the prevailing concept of circularity or more precisely "circularity rate" in relation to resource flows is critical. Eurostat has developed "*a single summary indicator of circularity of our economy at macroeconomic level. This indicator is called the 'circular material use rate' — referred to as the* **circularity rate** *— and* **it measures the contribution of recycled materials** *towards the overall use of materials*". Based on this approach the EU's circularity rate in 2023 is estimated to about 12%[7]. While the intent of having a "single summary" indicator for "circular material use rate" in the economy makes sense, however denoting it as "circularity rate" is confusing to the extent of misleading. In the context of circular economy, the term circularity has more comprehensive connotation which includes not only the mechanism of recycling for material recovery and reuse but also refers to mechanisms of maintenance, repair,

---

[6] *Waste to wealth, Peter Lacy and Jakob Rutqvist, 2015*
[7] *Circular economy- material flows (Eurostat webpage, accessed on March 12th, 2025)*

refurbishing and remanufacturing to enable life extension and reuse of assets (infrastructure, machinery, equipment, products, components etc.). Since the term "circularity rate" here merely represents the rate of recycled materials which is otherwise known simply as recycling it is hard to justify the necessity of using it as an alternative to recycling. Containing this term to such narrow meanings is not only misleading but also carries the risk of diminishing focus on far more important and preferred mechanisms for extending the life and continued use of assets, which are the core of circularity and circular economy frameworks. Use of "circularity rate" in similar meaning is adopted in other policy focused macroeconomic analyses such as the Circularity Gap Report (CGR) published annually by Circle Economy, a global impact organisation. According to the latest CGR published in 2024, the global circularity, that is the share of secondary materials consumed by the global economy, is 7.2%[8].

Setting aside concerns about the definition of circularity and its use as discussed above, for the moment, the following analysis of resource flows and associated circularity assumes the definition as it is, since this is how most of the material flow analyses are conducted in the context of circular economy. Primary purpose of the further analysis here is to understand how well the prevailing definition of circularity or circularity rate is representing the reality. In other words, it is to differentiate between "apparent circularity" and "real circularity". The analysis here makes use of a resource flow example from 2020 since published scientific works[9] provide detailed data for this year which is necessary for the intended purpose.

**Analysis:**

In 2020, resource input to the global economic system amounted to 104 Gt (billion metric tons), with about 40 Gt being the energetic and 64 Gt being the structural and technical materials. With share of about 15 Gt fossil fuels and 25Gt biomass, these flows generated about 45 Gt of air emissions and 25 Gt of solid and liquid waste. Further, 31 Gt of the total input ended up as net additions to the material stocks in the form of infrastructure, buildings, machinery, etc. It is important to note that about 9% of the materials were received via reverse flow of materials recovered through recycling. This example of resource flows can be very helpful in extending the analysis further for understanding circularity and circular economy as well as plausible transition or transformation scenarios.

Considering the total amount of recycled resources as a measure of circularity, as defined by Eurostat or in the circularity gap reports[10], the circularity rate or "circularity" turns out to be about 9% in this example. However, looking carefully on the numbers here, it is obvious that 39% of the resource input (40 Gt) being energetic is transformed to a consumed form that is not recoverable as original materials, if at all. This in turn means that in an idealised scenario of zero losses, which may occur due to diverse technical limitations, only 61% of the total resource input is available for a fully closed loop or circular economy. In other words, the 64 Gt of non-dissipative resource flows offer a circularity potential of maximum 61%, not more!  Keeping in view that dissipative materials are an inherent characteristic, especially of the fossil fuel driven economy, the figure of 64 Gt should be the reference for current calculations of circularity. Given the reverse flows of 9 Gt the given circularity metric thus improves from 9% to 14%.

Moving on, since 31 Gt out of the 64 Gt goes into the economy as stocks, locked in for several years and therefore not available for yearly recovery and reuse, the revised reference number then becomes 33 Gt. This makes the current circularity metric even better, pushing it to a whopping 27%. The picture suddenly looks much more optimistic in comparison to the reported 9% (8.6%)! In summarising this analysis, it is fair to conclude that circularity rate of 9% is valid only if all annual resource input to the economy is recoverable for reuse which is obviously not the case here. As we start deducting the non-recoverable proportion of the annual input, including the dissipative materials and the net addition to the stocks, we are left with the proportion of resources which today end up as solid and liquid waste but are assumed to be fully recoverable in an idealised case. In the given scenario, it is obvious that the real number on circularity is about 27% instead of the reported 9%, which should be considered only as apparent circularity. It is important to note that the role of non-recoverable materials (energetic and net addition to stocks) in lowering the circularity metric of the current economy has been emphasised is state of the art literature[11],

---

[8] *The Circularity Gap Report 2024 by Circle Economy Foundation*
[9] *Heinz Schandl, et al., Global material flows and resource productivity- The 2024 update, Journal of Industrial Ecology 2024;28:2012–2031*
[10] *The Circularity Gap Report 2020 by Circle Economy Foundation*
[11] *Willi Haas et al., How Circular is the Global Economy? An Assessment of Material Flows, Waste Production, and Recycling in the European Union and the World in 2005, Journal of Industrial Ecology 2015; Vol 19, No. 5*

however, this apparent circularity still remains the pivotal factor in mainstream analyses and discussions on circular economy.

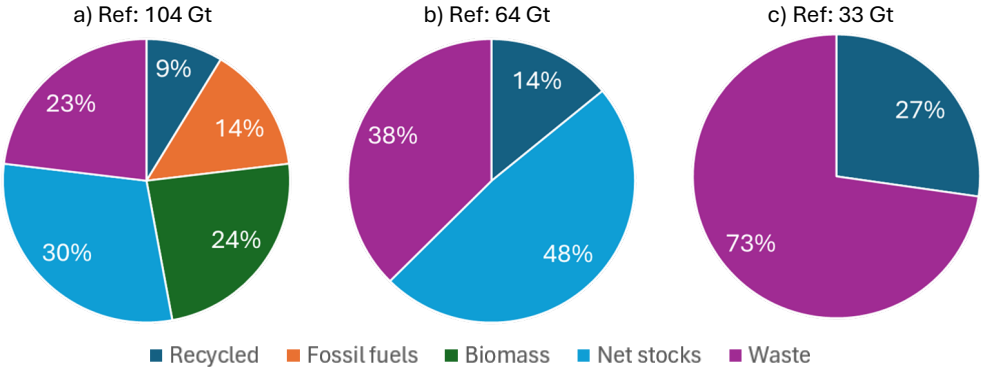

*Figure 1: Evolution of the circularity metric, from apparent 9% to real 27%, as the recoverable resource input to the economy reduces from 104 Gt to 33 Gt*

So far, this analysis has looked only at the material flows and assumed recycling for material recovery as the measure of circularity. While recycling as a measure of circularity is already a big question mark, an even more important question here is if there is one to one correspondence or proportionality between such circularity metric and economic value or the economy it creates. In other words, if we assume annual GDP as a measure of the economy, does this measure of circularity translate to economy of a proportional size as well. In other words, does the 9% or 27% circularity mean 9% or 27% of economy? And if not, how meaningful this measure is for understanding the impact of annual material flows on the economy that is envisioned to be transformed from linear to circular?

As highlighted in the beginning, several detailed studies on resource flow analyses including in relation to GDP have been carried out in the past years in the context of both linear and circular economy[12], however, understanding the relationship between the resource flows and stocks and the economic value associated with them remains challenging. It is also very difficult to find data that provides direct mapping of the economic value or the proportion of GDP that these flows and stocks create or are responsible for. The analysis here, however rudimentary it may be, seeks to fill this gap and acts as a springboard for debate and comprehensive analyses.

Since, total amount of recycled content as a measure of circularity is the reference here, it can be considered as is, for the economic analysis part as well. Assuming that primary sources of secondary materials are the global waste management activities and therefore can be considered sources for the 9 Gt of reverse flows here. A synthesis of the available information on economic value that the global waste management sector creates the most reliable number turns out to be $1.2 trillion[13]. Although this number represents economic value greater than created through the flow and use of recycled materials only, for the sake of simplification it is taken as is. The actual number representing direct value creation from the flows of recycled contents may be lower. Using this approximation and the circularity rate of about 9% in 2020, the proportion of GDP ($86 trillion in 2020) it generates, turns out to be 1.4%. If we instead chose to use the circularity rate of 14% or 27%, according to the advance analysis presented above, difference between the circular material flows and the economic value they create gets even higher. And with 14% or 27% circularity generating mere 1.4% of the of the GDP, the picture looks very grim. Is it surprising? Not at all! We understand that recycling, though an essential component of circularity, is the most energy intensive but the least value-preserving and therefore the last desirable loop for circularity and circular economy. The inner loops of repair, refurbish and remanufacture for reuse (continued use) of assets have significantly low energy requirements but are substantially more value-preserving and should therefore be an integral part of any circularity metric.

---

[12] *Fridolin Krausmann et al., Growth in global materials use, GDP and population during the 20th century, Ecological Economics 68 (2009) 2696–2705*
[13] *Global Waste Management Market Size & Outlook, 2023-2030 reports the global waste market contributing about $1.3 trillion to the global economy of about $85 trillion in 2020*

So far, we have examined only one type of flow, the reverse, and its relation to the economy which contributes little over 1% to the economy. It is then natural to wonder about the remaining 99% of GDP and its relation to the outstanding 91% of the forward flows. Even though 40% of the forward flows are of dissipative nature in their final fate, however, they are significant contributor to the GDP. Once again, numbers on direct correspondence between these 40% inputs and the value they create in the economy are hard to find, a reasonable estimate through data mining and calculations though is possible. Energy sector (fossil fuels), agriculture and food production, and chemicals and industrial materials are known to contribute annually $ 5-7 trillion, $4-5 trillion and $4-5 trillion[14], respectively. Approximating these numbers to a total of $15 trillion, a total of $16.2 trillion makes it about 19% of the annual GDP economy. With about 49% of the resource input so far, accounting only for 19% of the annual GDP, we are left with 51% of the resource input to account for the remaining 81% of the GDP!

As noted earlier, 25 Gt of the non-dissipative inputs end up as solid and liquid waste, which being unmanaged are most likely deteriorating than adding any value to the economy. Assuming a zero value-addition from this lost part of the inputs, we are left with 31 Gt of net addition to the stocks to account for about $69 trillion contribution to the GDP. In other words, about 30% of the resource input accounting for 81% of the economy! Sounds exciting, however it is not the whole truth, yet!

Net addition to stocks in 2020, captured as Net Fixed Capital Formation (NFCF) in the GDP can be estimated by using the rate of Gross Fixed Capital Formation (GFCF) and Consumption of Fixed Capital (CFC). Using the data available from several sources this number is estimated to be about 13% of the GDP[15]. With this information we can conclude that the 31 Gt of net addition to the stocks has contributed about $11 trillion to the GDP. Summing up all contributions from the resource inputs to the economy- $1.2 trillion from reverse flows, $15 trillion from dissipative flows and $11 trillion from net addition to the stocks- we reach to a total of $27 trillion which means that we are still missing $59 trillion from a total of $86 trillion annual GDP in 2020. Since we have considered all resource inputs regarding their estimated contribution to the GDP, we are left with the legacy stocks- already built societal infrastructure- to be the remaining contributor to the economy. In other words, about 69% of the economy in this case depended on legacy stocks, their use and management instead of the yearly resource inputs. Well obviously, the picture is not this black and white, since the overall economy is a mix of flows and stocks where the use and management of legacy stocks depend on energy supplies from dissipative resource flows, for example. Chemical and industrial materials sector is producing several non-dissipative outputs, such as long-life plastics, which end up being part of stocks in the economy. However, this analysis highlights clearly that contribution of annual resource inputs (flows and stocks) caters for less than one third of the GDP whereas the contribution from use and management of the legacy stocks dominates with more than two third of the GDP. This outcome is very much in line with the fact that about two third of the global GDP is generated by the services sector where the figure for 2020 was 65% and has been above 60% for the past two decades, at least[16].

---

[14] *Several different sources combined to extract these estimates, such as [Total revenue of the chemical industry worldwide from 2005 to 2022](#) from Staista.com*

[15] *Net Fixed Capital Formation (NFCF)= [Gross Fixed Capital Formation (global GFCF for year 2020 was 26% of GDP)](#)- Consumption of Fixed Capital (Global CFC average is estimated to be 13% based on several information sources since a single value is not available)=26-13=13%*

[16] [World Bank Group: Services, value added (% of GDP)- 1995 to 2022](#)

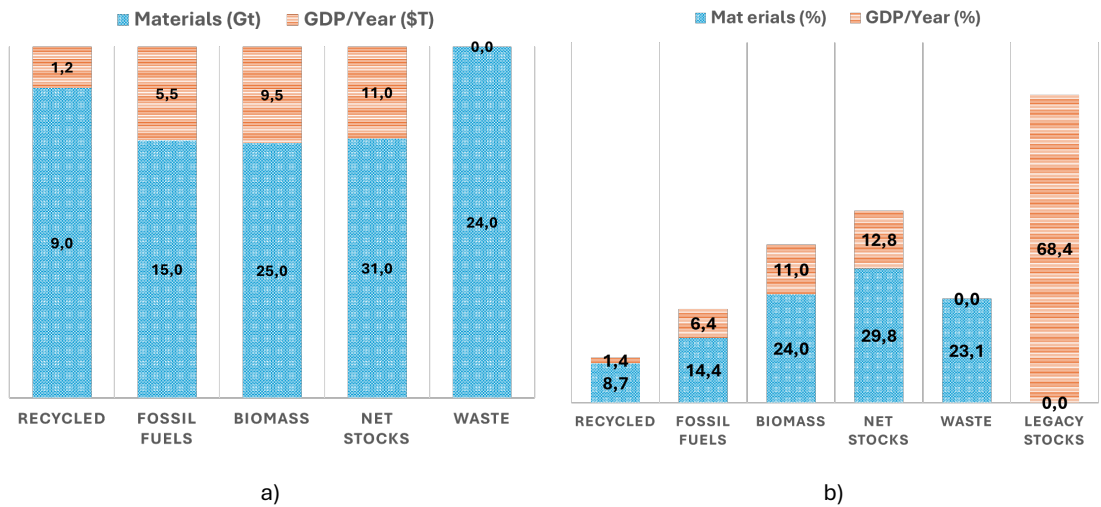

*Figure 2: Relationship between the categories of resource input and the economic value they create in terms of annual GDP*

**Discussion:**

As mentioned earlier, the studies on material flow analysis both in the context of linear and circular economy do emphasise on the impact of dissipative resources and net additions to stocks on circularity however the mainstream discussions remain focused on the low degree of circularity calculated based on total resource input including those which are irrecoverable for obvious technical or socioeconomic reasons. This confronts us with two challenges: first, that using recycling as a measure of circularity and not correcting it for the irrecoverable part of the resource input presents a skewed picture where the ambition of achieving 90 or 100% "circularity"[17] approaches to a level of impossibility. Second, maintaining recycling as the measure of circularity and correcting it for irrecoverable resources though enhances the figure on circularity from 9% to 27%, however, the fact that it contributes only about 1.4% to the economy (GDP here) makes the situation even worst since the target of achieving a circular economy in this scenario seems next to impossible. If all the recoverable material fraction could be converted to useful reverse flows (only one fourth was recovered in the 2020 material flow example), and assuming a proportional contribution to the GDP, their maximum potential will still be less than 6% of the GDP. Thus, assuming that we can attain the circularity rate of 100% (ideal), it will never mean reaching 100% circular economy. Therefore, using the mainstream "circularity rate", which is essentially the rate of recycling in the linear economy, as a measure of circularity is not only misleading but also detrimental to ambitions and efforts for transitioning to a circular economy. As the analysis here shows that the largest part of the economy even today depends on use and management of stocks, it is essential that circularity metrics include intensity of use (reuse) and management (maintain, repair, refurbish and remanufacture) as integral and prioritised part of the measure. Only then we can envision a circular economy!

Now, if we correct the measure of circularity in line with one of the core principles of circular economy that demands the use of resources at their highest added value for as long as feasible, the prevailing linear economy presents a very interesting picture. Analysis of sectoral contribution to economy shows that two third part of the GDP comes from services sector. The service sector economy is based primarily on the principle of use and management of societal stocks to ensure longer life for intended or improved services over time. In this scenario can we then say that we are already living in a "circular economy", however fraught with several short comings such as broken technical and bio circles which cause massive amount of resources ending as waste, create pollution and continuously deteriorate the natural eco system? For example, more than 23% of the total resource input goes to waste, which acts as a double-edged sword- zero contribution in the economy and serious damage to the natural eco system. It is obvious that an economy based on massively polluting emissions and waste cannot be called

---

[17] 100% "circularity" here means recovery of all materials that are recoverable using state of the art technical solutions and systems. For example, if we assume that all of 33 Gt can be recovered for reuse, we may claim to approach 100% circularity however never meaning to reach a fully circular economy

circular, however, we are surely living in an economy that is dominantly driven through core principles of circular economy. This realisation makes the target of achieving a fully circular economy way more feasible than understood from the perception that the world is moving in an opposite direction.

Keeping in view that two third of the economy (67%) can be run through use and management of legacy stocks, if we replace fossil fuels completely (ideal) with renewables, divert the waste to stock building where needed and enable recycling of all that is left over with the purpose of enhancing the use and management of stocks, a natural curiosity is if it is possible to stretch the figure of 67% to 100%? While an economy based entirely on use and management of legacy stocks "may be" possible however does this mean that the need for continuous input of virgin resources will be eliminated altogether? In a practical scenario which takes all technical and systemic limitations causing losses and inefficiencies in resource recovery processes and systems into account and is based on a growth driven economy, need for continuous resource input will persist. Even in a hypothetical scenario where all the limitations are overcome, the use and management of the stocks will still require continuous resource input regardless the economic model is based on growth or steady state (zero growth).

At this point, it is useful to present a perspective on dissipative part of the resources. It is important to note that most of the technical solutions envisioned to replace fossil fuel energy with renewable or clean energy are material intensive, however, they are not dissipative in the way fossil fuels are. Therefore, it is highly feasible that these solutions can be built as valuable stocks. Design of these solutions must take advantage of the circular economy principles ensuring their long-term use and management spanning over decades. Their end-of-life design should also make best use of the material circularity principle ensuring their efficient recovery and effective reuse thus reducing the need for constant input of dissipative virgin resources that are necessary otherwise.

Before concluding the discussion, it is critical to question if an economy driven by services sector is inherently resource conserving? This question obviously warrants a separate piece of research; however, a careful scan of the available information may help answer the question with necessary brevity.

Contribution of services sector in GDP of different economies varies significantly, ranging on average from 75% to 40% in high-income to low-income economies respectively[18]. While the material footprints of high-income countries where services sector contributes the highest to their economies remain significantly high, studies show that it is the consumption patterns that are responsible for their high material footprints and not the services per se. For example, in case of the European economy where the share of services sector is typically more than 70% of the GDP[19] and average material footprint is about 15 tonnes/capita[20], material intensity of services is the lowest out of six consumption domains including, housing, food, personal mobility, household goods, clothing and footwear. Services used the least materials per euro spent (< 0.25 kg/€ in 2021), which is roughly eleven times less than housing[21]. Combined with linear economy practices the exorbitant consumption patterns in the high-income economies, excessive material use in these economies is also driven by factors such as infrastructure needs, global supply chains resulting in imported goods with high embodied materials, high energy use for services and short product lifespans demanding frequent replacement of products (e.g., electronics, clothing). This scenario warrants that the work on circular economy must include the consumption aspect as an integral part of the equation. This has been one of the conclusive remarks of UNEP's report[22] as well. Negative impact of ignoring the consumption side of the equation is illustrated well on our urban roads daily where continuous production efficiency gains over last several decades appear nullified by hundreds of thousands of cars carrying around only one person in them, the driver!

While the consumption must be brought in as an integral part of the grand circular transition plan, the macroeconomic policies should continue pushing and provide support on the production side, for the development of maintenance, repair, refurbishing and remanufacturing businesses in societies across the globe. Scaling up these businesses is essential for economic feasibility and sustainability of circular solutions. This will not only spark innovations in manufacturing sectors and help OEMs to expedite their transition to circularity beyond recycling but will also strengthen local economies through numerous, small and medium scale businesses emerging in closer proximity where the OEM products are used and maintained. For example, manufacturing companies can combine the global forward supply chains with the local reverse supply chains if "remanufacturing-as-a-service" is developed as a new and prevalent business model. Similarly, "recycling-as-a service" can be developed as a widespread business model. Such developments can be accelerated through favourable policy measures including subsidies to establishing such businesses. With linear economy's fossil fuel sector getting subsidies of the order of $7 trillion[23], transition to circular economy is at sheer disadvantage let alone having a level playing field.

---

[18] *Trade in services for Development: Fostering sustainable growth and economic diversification-published by WTO and World Bank Group, 2023*
[19] *National accounts and GDP: EUROSTAT data , June 2024*
[20] *Europe's material footprint, European Environment Agency, Published 13 December 2024*
[21] *From data to decisions: material footprints in European policy making, European Environment Agency, Briefing published on 09 Oct 2024*
[22] *Bend the trend: Pathways to a liveable planet as resource use spikes, UNEP- Global Resources Outlook 2024*

**Conclusions**:

Mainstreaming recycling rate as the measure of circularity is not only misleading it is detrimental to a more comprehensive definition of circularity which prioritises the use and reuse of resources at their highest value added through mechanisms as maintenance, repair, refurbishing and remanufacturing. Secondly if circularity measure is disconnected from the contribution, it makes or can make to the GDP, its potential to transform the economy to a circular economy will not be clear and therefore the measures necessary for such transformation will be either misdirected or seriously fall short of achieving the objective. From resource intensity perspective, an economy is a consequence of dynamic interactions between resource flows and stocks. Keeping in view that use and management of stocks play a major role with dominant contribution to the economy, recycling alone will never be able to create a circular economy. Overemphasising the role of material flows and labelling their recovery as circularity is undermining the importance of stocks use and management. Circular transformation should focus on managing stocks and building new high-quality stocks where necessary. Minimising forward flows (virgin resources) through best use of the stocks and designing systems for best efficiency of the necessary reverse flows (secondary resources) should be primary objective of circular economy. The forward and reverse flows of materials should be subservient or a means to achieving this objective. It is therefore very important that the mainstream definition of circularity is corrected and its connection with economy is established and well understood for developing impactful strategies and policies.

---

[23] *Simon Black et al.,* IMF Fossil Fuel Subsidies Data: 2023 Update*, August 24, 2023*